\documentclass[aps,prl,twocolumn,showpacs,superscriptaddress,groupedaddress]{revtex4}  
\usepackage{graphicx}  
\usepackage{dcolumn}   
\usepackage{bm}        
\usepackage{amssymb}   
\usepackage{amsmath}
\usepackage{color}

\DeclareMathOperator{\sech}{sech}

\begin{document}


\title{Four-band insulator on a $\mathbb{Z}_2$ domain wall: an analytically solvable model for the interface between trivial and topological 2D insulators}

\def\ITA{Instituto Tecnol\'{o}gico de Aeron\'{a}utica, 12228-900 S\~{a}o Jos\'{e} dos Campos, SP, Brazil}
\def\JENA{Institut f\"ur Festk\"orpertheorie und -optik, Friedrich-Schiller-Universit\"at, 07743 Jena, Germany}

\author{F.L.~Freitas} \email{felipelopesfreitas@gmail.com} \affiliation{\ITA,\JENA}
\date{\today}

\begin{abstract}
A phenomenological model for the interface between trivial and topological two-dimensional insulators possessing the same band gap is presented. The model depends on three measurable parameters, the energy gap $E_g$, the Fermi velocity of the metallic edge states $v_F$ and the thickness of the interface $\Delta$ where the gap inversion occurs, and can be reduced to the Schr\"odinger equation for the modified P\"oschl-Teller potential, which admits an analytical solution. It is demonstrated that the underlying physics is determined by the adimensional parameter $\alpha=E_g\Delta/2\hbar v_F$, whose integral part determines the number of massive bound states at the interface. Furthermore, when $\alpha$ is exactly an integer, waves incident on the interface are never reflected. Results for parameters chosen in the typical scale of condensed matter systems are briefly discussed.
\end{abstract}

\maketitle


Topological insulators were brought to the forefront of theoretical physics research by the seminal work of Kane and Mele\cite{Kane2005a,Kane2005b}, where it was shown that two-dimensional periodic systems can have unusual physical properties due to the topology of their band structures, namely the quantization of Spin Hall conductivity, and the presence of spin-momentum locked gapless states at the border between insulators of distinct topological classes.

The metallic edge states can be understood in a variety of ways, such as the chiral zero mode first studied by Jackiw and Rebbi\cite{Jackiw1976,Hasan2010} or within the formalism of a four-band insulator model where the interaction between two distinct spin polarizations of the electrons put them in two different topological phases of Haldane's model\cite{Haldane1988} (see also Ref.~\onlinecite{Fruchart2013} and references therein), each giving rise to metallic edge states propagating in opposing directions at the border.

Despite the tremendous interest in two-dimensional topological insulators (TIs), the study of the physics at the interface of a TI and a regular insulator has been so far hampered by the lack of simple phenomenological models to describe this system. In this letter, I present such a model, inspired from field theory, valid when both the TI and the regular insulator have the same gap, and show that it can be solved analytically, deriving the full spectrum of midgap massive surface states, the chiral metallic states, and the reflection and transmission coefficients for scattering states.

Working with units $\hbar=v_F=1$, where $v_F$ is the Fermi velocity of the gapless states at the border, the Lagrangian we are interested in is, in 2+1 dimensions,

\begin{equation}
\mathcal{L} = \frac{1}{2}(\partial_\mu\phi)^2 - \frac{\lambda}{4}(\phi^2-\eta^2)^2 + i\bar{\psi}\gamma^\mu\partial_\mu\psi -g\phi\bar{\psi}\psi,
\end{equation}
where $\phi$ is a massless real scalar field, which acquires mass through spontaneous symmetry breaking, and $\psi$ is a four-component spinor, associated with the electron and hole states of different spin, which acquires a mass by coupling with the scalar field by means of the Yukawa term with coupling constant $g>0$.

We are interested in the fermionic states propagating through a fixed scalar field $\phi$. The self interaction term places the energy minimum at $\phi=\pm\eta$, and we look for static ($\partial_t\phi=0$) solutions, which are homogeneous in the $x$ coordinate ($\partial_x\phi=0$) and satisfy the boundary conditions $\phi(y=\pm\infty) = \pm\eta$.

Under these assumptions, the scalar and fermionic fields each acquire a mass at $|y|\to\infty$, given by $m_s=\sqrt{2\lambda}\eta$ and $m_f = g\eta$. Neglecting the Yukawa term, the equation for the scalar field has the well-known $\mathbb{Z}_2$ kink solution\cite{Vachaspati2006}

\begin{equation}
\phi = \eta\tanh(m_sy/2) = \eta\tanh(y/\Delta),
\end{equation}
where I have introduced the parameter $\Delta$, related to the thickness of the domain wall. The bound states for this system in 1+1 dimensions have been studied in Ref.~\onlinecite{Chu2008}, while scattering states have been considered in the limit $\Delta\to0$ in Ref.~\onlinecite{Campanelli2004}. In this work, I generalize both results to (2+1)D and also provide a more elementary solution to the problem.

The Dirac equation in the fixed scalar field background reads

\begin{equation}
(i\gamma^\mu\partial_\mu - g\phi)\psi = 0.
\end{equation}

To make the connection to topological insulators, we choose the Dirac matrices

\begin{equation}
\gamma^t = \left[
\begin{matrix}
\sigma_z & 0 \\
0 & \sigma_z
\end{matrix}
\right], \qquad
\gamma^x = \left[
\begin{matrix}
i\sigma_y & 0 \\
0 & -i\sigma_y
\end{matrix}
\right], \qquad
\gamma^y = \left[
\begin{matrix}
i\sigma_x & 0 \\
0 & i\sigma_x
\end{matrix}
\right],
\end{equation}
where $\sigma_i$ denote the Pauli spin matrices. The gamma matrices are block diagonal, and the upper and lower components represent the spin up and spin down bands, respectively. The equation reduces to two analogous equations for two-component spinors $\psi_1$ and $\psi_2$. We write the solution for these spinors as

\begin{equation}
\psi_i(x,y;t) = e^{i(q_xx-Et)}\left\{u_+(y)\left[
\begin{matrix}
\frac{1}{\sqrt{2}} \\
\frac{1}{\sqrt{2}}
\end{matrix}
\right]
+u_-(y)\left[
\begin{matrix}
\frac{-1}{\sqrt{2}} \\
\frac{1}{\sqrt{2}}
\end{matrix}
\right]\right\}.
\end{equation}

In the context of TIs, the quantity $q_x$ represents the $x$-component of the momentum around a Time-Reversal Invariant Momentum (TRIM)\cite{Fu2007}, given by $\mathbf{q}=\mathbf{k}-\pmb{\lambda}$, since it is the gap closing at this point that causes the appearance of metallic edge states\cite{Fruchart2013}. With the previous ansatz, we obtain

\begin{equation}
\left[
\begin{matrix}
E - g\phi & \mp q_x - \partial_y \\
\pm q_x -\partial_y & -E - g\phi
\end{matrix}
\right]\left[
\begin{matrix}
u_+ - u_- \\
u_+ + u_-
\end{matrix}
\right] = \left[
\begin{matrix}
0 \\ 0
\end{matrix}
\right],
\end{equation}
where the upper and lower signs appear in the spin up and down components, respectively. Rearranging the equations, we get

\begin{align}
(\partial_y + g\phi)u_+ &= -(E\pm q_x)u_-, \\
(\partial_y - g\phi)u_- &= (E\mp q_x)u_+. \label{eq:chiral}
\end{align}

Applying $(\partial_y\mp g\phi)$ on the left, the equations assume the same form for spin up and spin down:

\begin{equation}
-\partial_y^2 u_\pm + g(g\phi^2\mp\partial_y\phi)u_\pm = (E^2-q_x^2)u_\pm.
\label{eq:sch}
\end{equation}

From now on, we will work with the equation for $u_-$, as the one for $u_+$ can be solved with trivial modifications. Substituting $\phi$, defining $\alpha=2m_f/m_s$ and writing $q_y^2 = E^2-m_f^2-q_x^2$, we get a Schr\"odinger equation for the Modified P\"oschl-Teller potential\cite{Flugge1994}:

\begin{equation}
u'' +\left[q_y^2 +\frac{\alpha(\alpha-1)}{\Delta^2}\sech^2\left(\frac{y}{\Delta}\right)\right]u =0.
\end{equation}

It is well-known that, by performing the change of variables

\begin{equation}
z = \cosh^2(y/\Delta), \qquad u(z) = z^{\alpha/2} v(z),
\end{equation}
one obtains the hypergeometric equation

\begin{equation}
z(1-z)v'' + [c-(a+b+1)z]v'-abv = 0,
\end{equation}
where

\begin{equation}
a = \frac{1}{2}\left(\alpha+iq_y\Delta\right), \qquad b = \frac{1}{2}\left(\alpha-iq_y\Delta\right), \qquad c = \lambda + \frac{1}{2}.
\end{equation}

We want the solutions around the singular point $z=1$, given by

\begin{align}
_2&F_1(a,b,1+a+b-c;1-z), \\
(1-z)^{c-a-b} \ _2&F_1(c-a,c-b,1+c-a-b;1-z),
\end{align}
where the hypergeometric function $_2F_1$ is defined in terms of the Pochhammer symbol $(q)_n=\Gamma(q+n)/\Gamma(q)$ as

\begin{equation}
_2F_1(a,b,c;z) = \sum_{n=0}^\infty \frac{(a)_n(b)_n}{(c)_n} \frac{z^n}{n!}.
\end{equation}

These two solutions are even and odd, and we can write the wave function $u$ in terms of the spatial coordinate $y$ as

\begin{align}
u_A(y) &= \cosh^\alpha(y/\Delta)_2F_1(a,b,1/2;-\sinh^2y/\Delta), \\
u_B(y) &= \cosh^\alpha(y/\Delta)\sinh(y/\Delta) \times \\ \nonumber 
&_2F_1\left(a+\frac{1}{2},b+\frac{1}{2},\frac{3}{2},-\sinh^2(y/\Delta)\right).
\end{align}

To analyze the asymptotic behavior of the solutions at $|y|\to\infty$, we work with the hypergeometric function in the limit $z\to-\infty$ with the identity\cite{Abramowitz2012}

\begin{align}
_2F_1(a,b,c;&z\to-\infty) = (1-z)^{-a} \frac{\Gamma(c)\Gamma(b-a)}{\Gamma(b)\Gamma(c-a)} \nonumber \\ 
&+(1-z)^{-b}\frac{\Gamma(c)\Gamma(a-b)}{\Gamma(a)\Gamma(c-b)}
\end{align}

For scattering states, $q_y$ is real. Omitting unimportant normalization factors, we have

\begin{align}
u_A(y) &\to \left[ \frac{\Gamma(-iq_y\Delta)e^{iq_y(\Delta\ln 2 - |y|)}}{\Gamma\left(\frac{\alpha-iq_y\Delta}{2}\right)\Gamma\left(\frac{1-\alpha-iq_y\Delta}{2}\right)} + \mathrm{c.c.} \right]
\label{eq:asym1} \\
u_B(y) &\to \pm \left[ \frac{\Gamma(-iq_y\Delta)e^{iq_y(\Delta\ln 2 - |y|)}}{\Gamma\left(\frac{\alpha+1-iq_y\Delta}{2}\right)\Gamma\left(\frac{2-\alpha-iq_y\Delta}{2}\right)} + \mathrm{c.c.}\right] \label{eq:asym2}
\end{align}

Therefore, at $|y|\to\infty$, we have the asymptotic behavior

\begin{equation}
u_A(y) \to \cos(q_y|y|+\phi_A), \qquad u_B(y) \to \pm \cos(q_y|y|+\phi_B).
\end{equation}

To study scattering, we first notice that, since the fermion masses at $|y|\to\infty$ are the same, there is no refraction at the interface. The incoming wave either passes through or gets reflected. To understand the influence of spin, we need to compute the reflection and transmission matrices $R$ and $T$, seeking a solution of the form

\begin{equation}
\psi(y) = \left\{ 
\begin{array}{lr}
(e^{iq_yy} + Re^{-iq_yy})\psi_0, & y\to-\infty \\
Te^{iq_yy}\psi_0, & y \to+\infty
\end{array}
\right.
\end{equation}

We are about to see that the matrices $R$ and $T$ are proportional to the identity. For that, it suffices to study the scattering of the wave function given by $u_-(y)$. If we compose a solution of the form

\begin{equation}
u_- = Au_A + Bu_B,
\end{equation}
it is possible to show\cite{Flugge1994} that the reflection and transmission coefficients are given by
\begin{equation}
|R|^2 = \cos^2(\phi_A-\phi_B), \qquad |T|^2=\sin^2(\phi_A-\phi_B).
\label{eq:reftrans}
\end{equation}

The difference of the phases in the Gamma functions in \eqref{eq:asym1} and \eqref{eq:asym2} can be computed by applying $\Gamma(\bar{z})=\overline{\Gamma(z)}$, the Euler reflection formula

\begin{equation}
\Gamma(z)\Gamma(1-z) = \pi/\sin\pi z,
\end{equation}
and the identity

\begin{equation}
\arg\sin(a+bi) = \tan^{-1}\left(\cot a \tanh b\right).
\end{equation}

After elementary trigonometric manipulations, we get

\begin{equation}
\tan(\phi_A-\phi_B) = \frac{\sinh(\pi q_y\Delta)}{\sin(\alpha\pi)}.
\end{equation}

Substituting this value in \eqref{eq:reftrans}, we conclude that the scattering is periodic in $\alpha$, with period 1. Since our solution, valid for $u_-$, can be applied to $u_+$ by the transformation $\alpha\to\alpha+1$, both $u_-$ and $u_+$ waves scatter identically, and the reflection and transmission matrices should be proportional to the identity, as claimed before.

Besides that, another interesting physical phenomenon happens. When $\alpha=2m_f/m_s$ (or, in standard units, $\alpha=E_g\Delta/2\hbar v_F)$ is an integer, the wave is transmitted entirely ($|T|^2=1$) and the barrier is transparent. Meanwhile, when $\alpha$ is not an integer and the particle runs almost parallel to the wall (small $q_y$), $|R|^2\to 1$ and the interface is reflective.

To analyse the bound states, it suffices to set $q_y = i\kappa$, with $\kappa>0$. Under this condition, the parameters $a$ and $b$ become real, and the first terms in \eqref{eq:asym1} and \eqref{eq:asym2} grow like $e^{\kappa|y|}$. Therefore, the coefficients multiplying them must vanish in order to obtain a normalizable state. That happens when the arguments of the second Gamma function in the denominator are at the poles, located at nonpositive integers. Grouping all the conditions, we must have

\begin{equation}
E_n^2 - q_x^2 - m_F^2 = -\frac{(\alpha-1-n^-)^2}{\Delta^2} = -\frac{(\alpha-n^+)^2}{\Delta^2},
\end{equation}
and it follows that $n^+-n^- = 1$. The states with energy $E_n$ correspond to particle and hole states within the gap, with mass given by

\begin{equation}
m_n = m_f\sqrt{\frac{2n}{\alpha}-\frac{n^2}{\alpha^2}}, \qquad 1\leq n\leq \alpha
\end{equation}

Thus, the integral part of $\alpha$ gives the number of massive bound states trapped at the wall. The state with $n^+=0$ is special. The only way to satisfy \eqref{eq:sch} for $u_-$ is with $u_-=0$. Substituting this into \eqref{eq:chiral}, we obtain $E=q_x$ for spin-up and $E=-q_x$ for spin-down. These are, then, massless states, where the sign of the energy dependence gives the group velocity of the wave. It is readily seen that spin-up states propagate to the right, and spin-down ones to the left, as expected in TIs.

\begin{figure}[h!]
\includegraphics{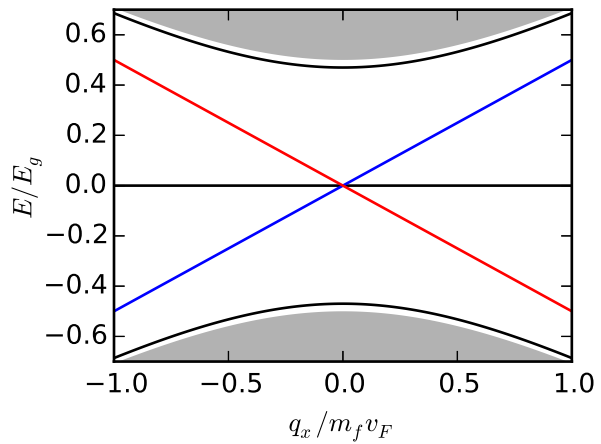}
\caption{\label{fig:disp} Dispersion relation for the system with parameters $v_F=c/300$, $\Delta=10$ nm and $E_g=0.2$ eV. The calculated value for $\alpha$ is approximately 1.52, giving rise to exactly one massive particle (and hole) state inside the gap.}
\end{figure}

To observe the massive bound states, it is necessary to have a material with a large band-gap, large wall thickness and small Fermi velocity. A plot with the full spectrum of states is presented in Fig.~\ref{fig:disp}, with parameters chosen in the typical range for condensed matter systems. It is readily seen that it should be possible to engineer materials so that $\alpha>1$.

In summary, I have presented a phenomenological model for the interface between trivial and topological 2D insulators with the same band gap $E_g$, that depends on the thickness of the interface $\Delta$ and the Fermi velocity $v_F$. The physics of the system is controlled by the adimensional parameter $\alpha=E_g\Delta/2\hbar v_F$, which controls the mass of the bound states trapped at the interface and also their number. I have shown that, when $\alpha$ is an integer, the interface is transparent to scattering states, and incoming particles are transmitted entirely, and also provided results for typical values of the measurable parameters. 

\begin{acknowledgements}
The author acknowledges financial support from CAPES (PVE grant no. 88887.116797/2016-00).
\end{acknowledgements}

\end{document}